\documentclass{ws-ijmpd}

\usepackage{graphicx}
\usepackage{epstopdf}
\begin{document}

\markboth{Bernardo Fraga, Carlos Arg\``{u}elles, Remo Ruffini}
{Fermions as Central Objects and Dark Halos in Galaxies}

%
\catchline{}{}{}{}{}
%

\title{SELF-GRAVITATING SYSTEM OF SEMIDEGENERATED FERMIONS AS CENTRAL OBJECTS AND DARK MATTER HALOS IN GALAXIES}

\author{Bernardo M. O. Fraga}

\address{ICRA and La Sapienza Universit\`a di Roma, Dipartimento di Fisica, P.le Aldo Moro 5\\
00185 Rome, Italy \\
Universit\'e de Nice Sophia-Antipolis, Parc Valrose \\
06108 Nice Cedex 2, France\\
bernardo.machado@icra.it}

\author{Carlos R. Arg\"uelles, Remo Ruffini}

\address{ICRA and La Sapienza Universit\`a di Roma, Dipartimento di Fisica, P.le Aldo Moro 5\\
00185 Rome, Italy\\
carlos.arguelles@icranet.org, ruffini@icra.it}



\maketitle

\begin{history}
\received{Day Month Year}
\revised{Day Month Year}
\comby{Managing Editor}
\end{history}

\begin{abstract}
We propose a unified model for dark matter haloes and central galactic objects as a self-gravitating system of semidegenerated fermions in thermal equilibrium. We consider spherical symmetry and then we solve the equations of gravitational equilibrium using the Fermi integrals in a dimensionless manner, obtaining the density profile and velocity curve. We also obtain scaling laws for the observables of the system and show that, for a wide range of our parameters, our model is consistent with the so called universality of the surface density of dark matter. 
\end{abstract}


\section{Introduction}	

Despite being discovered 75 years ago\cite{zwicky}, it still not well known what Dark Matter (DM) is composed of, and its properties. The current paradigm is that DM is composed of slowly moving very massive particles, thus being called Cold Dark Matter\cite{blumenthal} (CDM). Despite its high predictive power and being in good agreement with large scale structure observations, there is still some controversy respecting the small scale structure formation. There are a lot of different DM profiles, some empirical\cite{burkert,einasto} and some based on N-body simulations\cite{nfw1,nfw2}, with different properties. Although CDM is the most accepted model, some problems still remain giving it the place to another type of DM, the Warm Dark Matter (WDM), with masses in an intermediate range between some KeVs to some MeVs. It can solve some of the problems of CDM, but also has some of its own. So, despite more than 50 years of research in DM, there is still a lot to explore. 
\par On a complete unrelated field, it is believed today that there is a supermassive black hole on the center of almost every galaxy. Recent observations\cite{wmaphaze} of a microwave emission from the center of the Milky Way suggest that there is a considerable amount of DM in the center. Given these recent results, it could be possible that there is no black hole, but instead a distribution of DM particles. Some authors\cite{fermionball} suggested that an extended distribution of fermions, a "fermion ball", supported by degeneracy pressure, is present on the center of galaxies instead of a supermassive black hole.
\par We propose here a preliminary model for DM that tries to explain both the DM halos and the objects at the centers of galaxies by a self-gravitating system of semidegenerated fermions.
 
\section{Model}
The equilibrium configurations of a self-gravitating semi-degenerate system of fermions were already studied by others\cite{rr}. We are considering spherical symmetry, so the spacetime is described by the Schwarzchild metric
\begin{equation}
ds^2=-e^{\nu(r)}dt^2+e^{\lambda(r)}dr^2+r^2(d\theta^2+\sin^2\theta d\phi^2).
\end{equation}
Solving the Einstein equation we get the equilibrium equations:
\begin{eqnarray}
\frac{dP}{dr}&=&-\frac{G}{c^2}\frac{(P+\rho c^2)(M_r+4\pi\rho r^3)}{r(rc^2-2GM)} \label{dpdr} \\
\frac{dM_r}{dr}&=&4\pi\rho r^2 \label{dmdr},
\end{eqnarray}
where $M_r$ the mass within a radius $r$, $G$ the gravitational constant, $c$ the speed of light and $\rho$ and $P$ the mass-energy density and the pressure respectively, given by
\begin{eqnarray}
\rho&=&m\frac{g}{h^3}\int_0^{\infty} \left(1+\frac{\epsilon}{mc^2}\right)\frac{1}{e^{(\epsilon-\mu)/kT}+1}d^3p \label{rho} \\
P&=&\frac{2g}{3h^3}\int_0^{\infty}\left(1+\frac{\epsilon}{2mc^2}\right)\left(1+\frac{\epsilon}{mc^2}\right)^{-1}\frac{\epsilon}{e^{(\epsilon-\mu)/kT}+1} d^3p, \label{press}
\end{eqnarray}
where $g=2s+1$ is the multiplicity of states, $h$ is the Planck constant, $m$ is the rest mass of the particle, $k$ the Boltzmann constant, $T$ the temperature, $\mu$ the chemical potential and we write the integrals in terms of the kinetic energy of a single particle $\epsilon=\sqrt{p^2c^2+m^2c^4}-mc^2$. 
\par To integrate numerically the set of equations, it is useful to change our variables to dimensionless ones:
\begin{equation}
\rho=\frac{c^2}{G\chi^2}\hat{\rho}, \hspace{0.3cm} P=\frac{c^4}{G\chi^2}\hat{P}, \hspace{0.3cm} M_r=\frac{c^2\chi}{G}\hat{M}_r, \hspace{.3cm} r=\chi\hat{r},
\end{equation}
where 
\begin{equation}
\chi=\frac{\hbar}{mc}\frac{m_p}{m}\left(\frac{8\pi^3}{g}\right)^{1/2}
\end{equation}
has dimension of a length and $m_p=(\hbar c/G)^{1/2}$ is the Planck mass. Also, we write Eq. \ref{dpdr} in terms of the degeneracy parameter $\theta=\mu/kT$ (see Refs. \refcite{rr,klein} for details). The final set of dimensionless equations is shown in Ref. \refcite{rr}, which we integrate with initial conditions $M(0)=0$ and $\theta(0)=\theta_0$.

\par It is important to notice that we can solve the whole system completely in a dimensionless way without specifying a mass for the particle, which appears only in the characteristic length $\chi$ and in the definition of the temperature parameter $\beta_0$; however, this parameter can be uniquely determined by the surface velocity, as will be shown later. Thus, the mass remains as only a scaling parameter and we are left with two free parameters, $\beta_0$ and $\theta_0$.

\subsection{Properties of semidegenerate configurations}
\begin{figure}[!ht]
\centering
\includegraphics[width=0.45\linewidth]{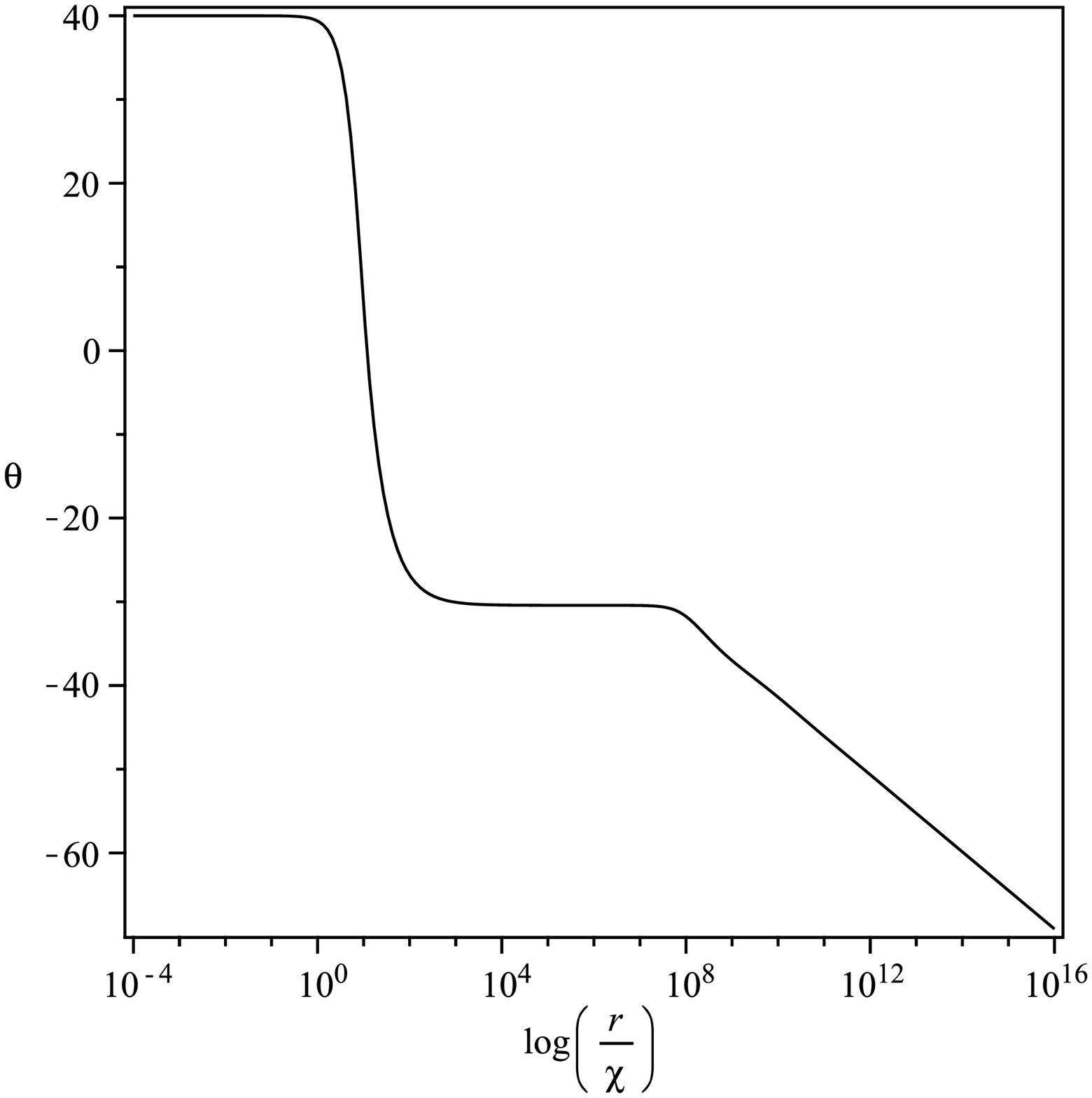}
\includegraphics[width=.45\linewidth]{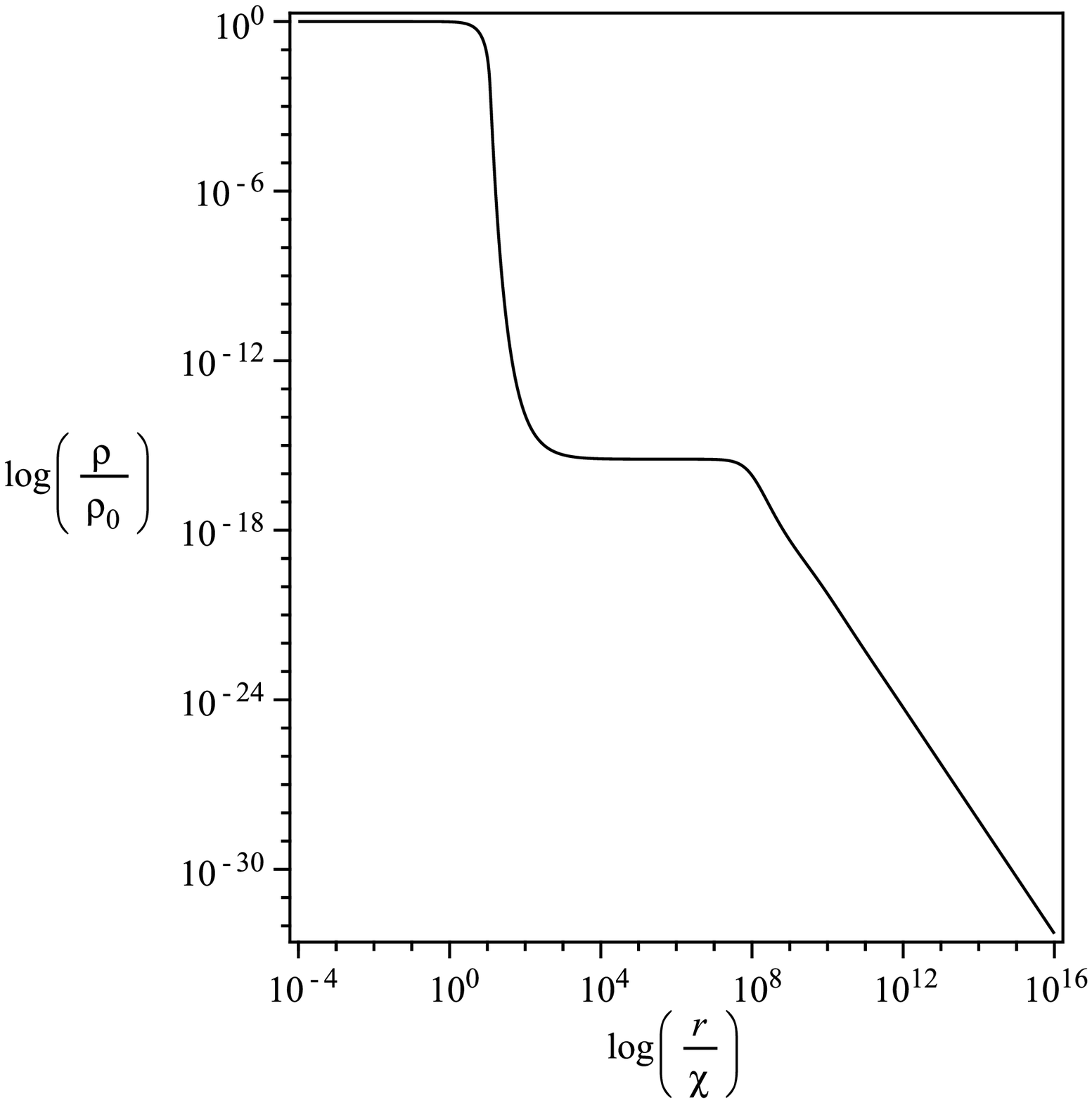}
\caption{Degeneracy parameter and density profile (left and right) for $\theta_0=40$ and $\beta_0=10^{-8}$.}

\label{fig1}
\end{figure}

Galactic halos are thought to be composed of Cold Dark Matter (CDM), so that $\beta_0\ll1$.Also, since we are interested in semidegenerate configurations, $ \theta\ $ must be a non negative parameter\cite{rr}. In Fig. \ref{fig1} we plot the degeneracy parameter and the density profile of the system as a function of the rescaled coordinate $\hat{r}$.

\par As we can see from these plots, there are three different regions:
\begin{itemize}
\item A degenerate core of almost constant density;
\item A "\textit{plateau}", or inner halo, also with almost constant density
\item The outer halo, where the density scales with $r^{-2}$.
\end{itemize}
Since the tail of the distribution scales like that, the system is unbounded and with infinite mass. In reality, the system will be bound by tidal interactions, and a cutoff in the distribution function must be used\cite{rrcutoff}. The astrophysically relevant quantity is the rotation velocity, given by (non-relativistic regime, since $\beta\ll1$):
\begin{equation}
v=\sqrt{\frac{GM_r}{r}}=c\sqrt{\frac{\hat{M}_r}{\hat{r}}}.
\end{equation}
\begin{center}

\begin{figure}[!ht]
\centering
\includegraphics[width=0.5\linewidth]{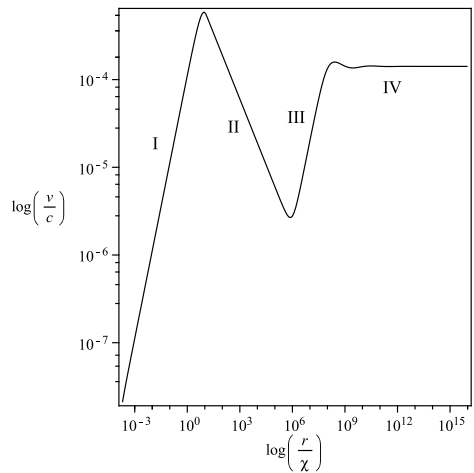}
\caption{Rotation curve for $\theta_0=40$ and $\beta_0=10^{-8}$.}
\label{fig2}
\end{figure}
\end{center}
Also in the rotation curve (Fig. \ref{fig2}) we can see four different regions:
\begin{itemize}
\item I: The degenerate core, where $v\propto r$;
\item II: The first part of the inner halo, where the mass of the core prevails over the halo and $v\propto r^{-1/2}$;
\item III: Second part of the inner halo, where the mass of the inner halo prevails over the halo, and $v\propto r$;
\item IV: The outer halo, where the velocity tends to constant after some oscillations.
\end{itemize}
\par This behavior is exactly what is observed for the outer part of the curve.
\subsection{Scaling laws}
To confront the model with observations, it is useful to see how the physical quantities scale with the parameters of our model. Our aim is to describe the galactic center and the dark matter halo, so we need to define both masses and radius for these regions. 
\par The radius of the core ($r_c$) is defined as the radius for the maximum velocity in region I; the mass is then $M_c(r_c)$; for the halo ($r_h$), we take the radius where the velocity is maximum in region III, with corresponding mass $M_h(r_h)$; finally, the last observable is the surface velocity ($v_{\infty}$), i.e., the velocity in the outer halo where it is constant. The scalings for these parameters are 
\begin{eqnarray}
\log\frac{M_c}{M_{\odot}}&=&11.732+0.75\log\beta_0+0.75\log\theta_0-2\log\frac{m}{\rm{keV}} \label{Mc}\\
\log\frac{r_c}{\rm{pc}}&=&-1.728-0.25\log\beta_0-0.25\log\theta_0-2\log\frac{m}{\rm{keV}} \label{rc}\\
\log\frac{M_h}{M_{\odot}}&=&13.646+0.75\log\beta_0+0.16\theta_0-2\log\frac{m}{\rm{keV}} \label{Mh}\\
\log\frac{r_h}{\rm{pc}}&=&-0.225-0.25\log\beta_0+0.16\theta_0-2\log\frac{m}{\rm{keV}} \label{rh}\\
\log\frac{v_{\infty}}{\rm{km/s}}&=&5.63+0.5\log\beta_0 \label{vinf}
\end{eqnarray}
\par As we can see, the temperature parameter at the center, $\beta_0$, can be uniquely determined by the observed surface velocity of the galaxy. Another interesting feature is that, while for the core the mass and radius scale with $\theta_0^{\alpha}$, in the halo they scale with $\alpha^{\theta_0}$. These scaling laws are valid for $\theta_0\in[20,110]$ and $\log\beta_0\in[-10,-5]$.

\subsubsection{Applications of the model}
We can apply this model, for example, to the Milky Way. The observed halo velocity is $v_{\infty}\approx200$ km/s, which gives $\beta_0=0.22\times10^{-6}$. Taking the observations of the Milky way, $M_c\approx4\times10^6~ M_{\odot}$, $M_h=2\times10^{11}~ M_{\odot}$ and $r_h=12$ kpc; and chosing Eqs. \ref{vinf},\ref{Mc} and \ref{rh} for the three unknown parameters we have:
\begin{eqnarray}
m&=&13~\rm{keV} \\
\theta_0&=&33,
\end{eqnarray}
which gives us $r_c\approx10^{-2}$ pc. This is unfortunately larger than the apocenter of the closest star to the center, ie. the star S2. 
\par Recently, some authors\cite{universality} found that, by fitting velocity curves with a Burkert profile, the dark matter surface density is constant within one dark matter halo scale-length for a wide span of galaxy magnitudes, $\rho_0r_0=141~M_{\odot}\rm{pc}^{-2}$. This leads to a maximal acceleration due to DM at $r_0$, $a_{DM}(r_0)=3.2\times10^{-9}~ \rm{cm/s^2}$. Using this value in our model, taking into account that $a_{DM}(r_h)=\frac{GM_h}{r_h^2}$, we get from Eqs. \ref{Mh} and \ref{rh}
\begin{equation}
1.27\log\beta_0-0.16\theta_0+2\log\frac{m}{\rm{keV}}=-11.7362
\end{equation}
\par Common halo velocities range from tens to hundreds km/s, ie. $\log\beta_0\in[-7.9,-6.3]$ from Eq. \ref{vinf}, and using the DM fermion mass obtained in the earlier application, we get that $\theta_0\in[23,38]$. So, for a wide range of the parameters with a fixed DM mass, our model is in good agreement with this universality law.
\par Despite not being able to describe the core and the DM halo of our galaxy, our model is in good agreement with some observations. Of course, a cutoff in the distribution function is needed to avoid having an infinite mass, but the aim of this preliminary work is to show the scope of the fully relativistic treatment of the model in the application to galactic haloes and also to our Milky Way, concluding that some new physics must be added to the model in order to have a unified approach for the later case.



\begin{thebibliography}{00}  

\bibitem{zwicky}
F. Zwicky, {\it Helvetica Phys. Acta} {\bf 6} (1933) 110

\bibitem{blumenthal}
G. R. Blumenthal, S. M. Faber, J. R. Primack and M. J. Rees, {\it Nature} {\bf 311} (1984) 517

\bibitem{burkert}
A. Burkert, {\it ApJL} {\bf 447} (1995) L25

\bibitem{einasto}
J. Einasto {\it Astrofizika} {\bf 5} (1969) 137

\bibitem{nfw1}
J. F. Navarro, C. S. Frenk, and D. M. White {\it ApJ} {\bf 462} (1996) 563-575

\bibitem{nfw2}
J. F. Navarro, C. S. Frenk, and D. M. White {\it ApJ} {\bf 490} (1997) 493-508

\bibitem{wmaphaze}
D. Hooper, D. P. Finkbeiner and G. Dobler, {\it Phys. Rev. D} {\bf 76} (2007) 083012

\bibitem{fermionball}
D. Tsiklauri and R. D. Viollier, {\it ApJ} {\bf 500} (1998) 591

\bibitem{rr} 
J.G. Gao, M. Merafina and R. Ruffini, {\it A\&A} {\bf 235} (1990) 1.

\bibitem{klein}
O. Klein, {\it Rev. Mod. Phys.} {\bf 21} (1949) 531

\bibitem{rrcutoff}
G. Ingrosso, M. Merafina, R. Ruffini and F. Strafella, {\it A\&A} {\bf 258} (1992) 223

\bibitem{universality}
G. Gentile, B. Famaey, H. Zhao and P. Salucci, {\it Nature} {\bf 461} (2009) 7264

\end{thebibliography}
\end{document}